\documentclass[12pt]{article}

\catcode`\@=11
\def\@citex[#1]#2{%
\if@filesw \immediate \write \@auxout {\string \citation {#2}}\fi
\@tempcntb\m@ne \let\@h@ld\relax \def\@citea{}%
\@cite{%
  \@for \@citeb:=#2\do {%
    \@ifundefined {b@\@citeb}%
      {\@h@ld\@citea\@tempcntb\m@ne{\bf ?}%
      \@warning {Citation `\@citeb ' on page \thepage \space undefined}}%
      {\@tempcnta\@tempcntb \advance\@tempcnta\@ne%
      \@tempcntb\number\csname b@\@citeb \endcsname \relax%
      \ifnum\@tempcnta=\@tempcntb 
        \ifx\@h@ld\relax%
          \edef \@h@ld{\@citea\csname b@\@citeb\endcsname}%
        \else%
          \edef\@h@ld{\ifmmode{-}\else--\fi\csname b@\@citeb\endcsname}%
        \fi%
      \else
        \@h@ld\@citea\csname b@\@citeb \endcsname%
        \let\@h@ld\relax%
      \fi}%
    \def\@citea{,\penalty\@highpenalty\,}%
  }\@h@ld
}{#1}}

\def\@citeb#1#2{{[#1]\if@tempswa , #2\fi}}
\def\@citeu#1#2{{$^{#1}$\if@tempswa , #2\fi }}
\def\@citep#1#2{{#1\if@tempswa , #2\fi}}

\def\bcites{         
        \catcode`\@=11
        \let\@cite=\@citeb
        \catcode`\@=12
}

\def\upcites{         
        \catcode`\@=11
        \let\@cite=\@citeu
        \catcode`\@=12
}

\def\plaincites{      
        \catcode`\@=11
        \let\@cite=\@citep
        \catcode`\@=12
}

\newcount\hour
\newcount\minute
\newtoks\amorpm
\hour=\time\divide\hour by 60
\minute=\time{\multiply\hour by 60 \global\advance\minute by-\hour}
\edef\standardtime{{\ifnum\hour<12 \global\amorpm={am}%
        \else\global\amorpm={pm}\advance\hour by-12 \fi
        \ifnum\hour=0 \hour=12 \fi
        \number\hour:\ifnum\minute<10 0\fi\number\minute\the\amorpm}}
\edef\militarytime{\number\hour:\ifnum\minute<10 0\fi\number\minute}

\def\draftlabel#1{{\@bsphack\if@filesw {\let\thepage\relax
   \xdef\@gtempa{\write\@auxout{\string
      \newlabel{#1}{{\@currentlabel}{\thepage}}}}}\@gtempa
   \if@nobreak \ifvmode\nobreak\fi\fi\fi\@esphack}
        \gdef\@eqnlabel{#1}}
\def\@eqnlabel{}
\def\@vacuum{}
\def\marginnote#1{}
\def\draftmarginnote#1{\marginpar{\raggedright\scriptsize\tt#1}}
\def\draft{
        \pagestyle{plain}
        \overfullrule=2pt
        \oddsidemargin -.5truein
        \def\@oddhead{\sl \phantom{\today\quad\militarytime} \hfil
        \smash{\Large\sl DRAFT} \hfil \today\quad\militarytime}
        \let\@evenhead\@oddhead
        \let\label=\draftlabel
        \let\marginnote=\draftmarginnote
        \def\ps@empty{\let\@mkboth\@gobbletwo
        \def\@oddfoot{\hfil \smash{\Large\sl DRAFT} \hfil}
        \let\@evenfoot\@oddhead}
        \def\@eqnnum{(\theequation)\rlap{\kern\marginparsep\tt\@eqnlabel}%
        \global\let\@eqnlabel\@vacuum}  }

\def\blackfonts{
        \font\blackboard=msbm10 scaled\magstep1
        \font\blackboards=msbm8
        \font\blackboardss=msbm6
}

\def\prep{         
        \catcode`\@=11
        \input art10.sty
        \catcode`\@=12
        
        \let\small\null
        \def\blackfonts{
                \font\blackboard=msbm10
                \font\blackboards=msbm7
                \font\blackboardss=msbm5
        }
        \let\sl\it
        \twocolumn
        \sloppy
        \voffset=-2.54truecm
        \hoffset=-2.54truecm
        \flushbottom
        \parindent 1em
        \leftmargini 2em
        \leftmarginv .5em
        \leftmarginvi .5em
        \marginparwidth 48pt
        \marginparsep 10pt
        \setlength{\columnsep}{2truecm}
        \setlength{\textwidth}{25.4truecm}
        \setlength{\textheight}{17truecm}
        \baselineskip=16pt
        \oddsidemargin .18truein
        \evensidemargin .17truein
}

\def\eqalign#1{\null\,\vcenter{\openup\jot\m@th
  \ialign{\strut\hfil$\displaystyle{##}$&$\displaystyle{{}##}$\hfil
      \crcr#1\crcr}}\,}
\def\eqalignno#1{\displ@y \tabskip\centering
  \halign to\displaywidth{\hfil$\@lign\displaystyle{##}$\tabskip\z@skip
    &$\@lign\displaystyle{{}##}$\hfil\tabskip\centering
    &\llap{$\@lign##$}\tabskip\z@skip\crcr
    #1\crcr}}

\def\section{\@startsection {section}{1}{\z@}{3.ex plus 1ex minus
 .2ex}{2.ex plus .2ex}{\large\bf}}
\def\subsection{\@startsection{subsection}{2}{\z@}{2.75ex plus 1ex minus
 .2ex}{1.5ex plus .2ex}{\bf}}        

\def\appendix{{\newpage\section*{Appendix}}\let\appendix\section%
        {\setcounter{section}{0}
        \gdef\thesection{\Alph{section}}}\section}
\def\thefootnote{\arabic{footnote}}
\def\abstract{\if@twocolumn
\section*{Abstract}
\else 
\begin{center}
{\bf Abstract\vspace{-.5em}\vspace{0pt}}
\end{center}
\quotation
\fi}

\catcode`\@=12

\def\d{\partial}

\def\sqr#1#2{{\vcenter{\vbox{\hrule height.#2pt\hbox{\vrule width.#2pt 
height#1pt \kern#1pt \vrule width.#2pt}\hrule height.#2pt}}}}

\def\=d{\,{\buildrel\rm def\over =}\,}

\def\S{\hbox{\bbf S}}

\def\i3p{\p32\int d^3p}

\def\As{A\hbox to 1pt{\hss /}}
\def\np4{\int d^4p_1\cdots d^4p_{n-1}\, }

\def\nx4{\int d^4x_1\ldots d^4x_n\, }

\def\kon#1#2{\vbox{\halign{##&&##\cr
\lower4pt\hbox{$\scriptscriptstyle\vert$}\hrulefill &
\hrulefill\lower4pt\hbox{$\scriptscriptstyle\vert$}\cr $#1$&
$#2$\cr}}}

\def\konv#1#2#3{\hbox{\vrule height12pt depth-1pt}
\vbox{\hrule height12pt width#1cm depth-11.6pt}
\hbox{\vrule height6.5pt depth-0.5pt}
\vbox{\hrule height11pt width#2cm depth-10.6pt\kern5pt
      \hrule height6.5pt width#2cm depth-6.1pt}
\hbox{\vrule height12pt depth-1pt}
\vbox{\hrule height6.5pt width#3cm depth-6.1pt}
\hbox{\vrule height6.5pt depth-0.5pt}}
\def\konu#1#2#3{\hbox{\vrule height12pt depth-1pt}
\vbox{\hrule height1pt width#1cm depth-0.6pt}
\hbox{\vrule height12pt depth-6.5pt}
\vbox{\hrule height6pt width#2cm depth-5.6pt\kern5pt
      \hrule height1pt width#2cm depth-0.6pt}
\hbox{\vrule height12pt depth-6.5pt}
\vbox{\hrule height1pt width#3cm depth-0.6pt}
\hbox{\vrule height12pt depth-1pt}}

\def\konw#1#2#3{\hbox{\vrule height12pt depth-1pt}
\vbox{\hrule height12pt width#1cm depth-11.6pt}
\hbox{\vrule height6.5pt depth-0.5pt}
\vbox{\hrule height12pt width#2cm depth-11.6pt \kern5pt
      \hrule height6.5pt width#2cm depth-6.1pt}
\hbox{\vrule height6.5pt depth-0.5pt}
\vbox{\hrule height12pt width#3cm depth-11.6pt}
\hbox{\vrule height12pt depth-1pt}}

\def\i{{\rm int}}

\def\a{{\rm av}}

\def\m3{{\mu_1\mu_2\mu_3}}

\def\p{{(+)}}

\def\be{\begin{equation}}       \def\eq{\begin{equation}}
\def\ee{\end{equation}}         \def\eqe{\end{equation}}

\def\bea{\begin{eqnarray}}      \def\eqa{\begin{eqnarray}}
\def\ena{\end{eqnarray}}        \def\eea{\end{eqnarray}}
                                \def\eqae{\end{eqnarray}}

\def\ba{\begin{array}}
\def\ea{\end{array}}
\def\unit{1 \hskip-.3em \raise2pt\hbox{$ \scriptstyle |$ } }

\def\a{\alpha}
\def\b{\beta}
 
\def\d{\delta}

\def\i{\iota}

\def\m{\mu}
\def\n{\nu}
  
\def\p{\pi}                
\def\t{\tau}

\def\G{\Gamma}

\def\L{\Lambda}

\def\S{\Sigma}

\def\ci{{\cal I}}

\def\bop#1{\setbox0=\hbox{$#1M$}\mkern1.5mu
        \vbox{\hrule height0pt depth.04\ht0
        \hbox{\vrule width.04\ht0 height.9\ht0 \kern.9\ht0
        \vrule width.04\ht0}\hrule height.04\ht0}\mkern1.5mu}
\def\pa{\partial}                              

\def\>{\rangle} 

\def\<{\langle} 
\def\Dsl{D \hskip-.6em \raise1pt\hbox{$ / $ } }

\def\sl#1{\rlap{\hbox{$\mskip 1 mu /$}}#1}
\def\leftrightarrowfill{$\mathsurround=0pt \mathord\leftarrow \mkern-6mu
       \cleaders\hbox{$\mkern-2mu \mathord- \mkern-2mu$}\hfill
       \mkern-6mu \mathord\rightarrow$}
\def\dvec#1{\vbox{\ialign{##\crcr         
       \leftrightarrowfill\crcr\noalign{\kern-1pt\nointerlineskip}
       $\hfil\displaystyle{#1}\hfil$\crcr}}}          
\def\hook#1{{\vrule height#1pt width0.4pt depth0pt}}
\def\leftrighthookfill#1{$\mathsurround=0pt \mathord\hook#1
       \hrulefill\mathord\hook#1$}
\def\underhook#1{\vtop{\ialign{##\crcr                 
       $\hfil\displaystyle{#1}\hfil$\crcr
       \noalign{\kern-1pt\nointerlineskip\vskip2pt}
       \leftrighthookfill5\crcr}}}
\def\smallunderhook#1{\vtop{\ialign{##\crcr      
       $\hfil\scriptstyle{#1}\hfil$\crcr
       \noalign{\kern-1pt\nointerlineskip\vskip2pt}
       \leftrighthookfill3\crcr}}}

\def\sfrac#1#2{{\vphantom1\smash{\lower.5ex\hbox{\small$#1$}}\over
       \vphantom1\smash{\raise.4ex\hbox{\small$#2$}}}} 
\def\bfrac#1#2{{\vphantom1\smash{\lower.5ex\hbox{$#1$}}\over
       \vphantom1\smash{\raise.3ex\hbox{$#2$}}}}      
\def\afrac#1#2{{\vphantom1\smash{\lower.5ex\hbox{$#1$}}\over#2}}  
\def\on#1#2{{\buildrel{\mkern2.5mu#1\mkern-2.5mu}\over{#2}}}
\def\ddt#1{\on{\hbox{\LARGE .\kern-2pt.}}#1}             
\def\tdt#1{\on{\hbox{\LARGE .\kern-2pt.\kern-2pt.}}#1}   

\def\boxes#1{
       \newcount\num
       \num=1
       \newdimen\downsy
       \downsy=-1.5ex
       \mskip-2.8mu
       \bo
       \loop
       \ifnum\num<#1
       \llap{\raise\num\downsy\hbox{$\bo$}}
       \advance\num by1
       \repeat}
\def\boxup#1#2{\newcount\numup
       \numup=#1
       \advance\numup by-1
       \newdimen\upsy
       \upsy=.75ex
       \mskip2.8mu
       \raise\numup\upsy\hbox{$#2$}}

\newskip\humongous \humongous=0pt plus 1000pt minus 1000pt
\def\caja{\mathsurround=0pt}
\def\eqalign#1{\,\vcenter{\openup2\jot \caja
       \ialign{\strut \hfil$\displaystyle{##}$&$
       \displaystyle{{}##}$\hfil\crcr#1\crcr}}\,}
\newif\ifdtup

\def\to{\rightarrow}

\def\1ov4{{1\over 4}}

\def\pa{\partial}

\def\ddt{\dot{\t}}

\def\pa{\partial}

\def \foot{\footnote}

\def \const {{\rm const}}

\def\pa{\partial}
\renewcommand{\a}{\alpha}
\renewcommand{\b}{\beta}

\renewcommand{\d}{\delta}

\newcommand{\beq}{\begin{equation}}
\newcommand{\eeq}{\end{equation}}

\def\ba{\begin{eqnarray}}
\def\ea{\end{eqnarray}}

\begin{document}


\null\vskip-24pt \hfill AEI-2000-078 \vskip-10pt\hfill
OHSTPY-HEP-T-00-032 \vskip-10pt 
\hfill {\tt hep-th/0012080} \vskip0.2truecm
\begin{center}
\vskip 0.2truecm {\Large\bf
Tachyon condensation and universality of DBI action
}\\
\vskip 0.5truecm
{\bf G. Arutyunov$^{*,**}$ \footnote{emails: {\tt
agleb, theisen @aei-potsdam.mpg.de;\ \  frolov,
 tseytlin @mps.ohio-state.edu
}
\newline
$^{**}$On leave of absence from Steklov Mathematical Institute,
Gubkin str. 8,
117966, Moscow, Russia}, 
S. Frolov$^{\dagger ,**}$, 
S. Theisen$^*$ 
and A.A. Tseytlin$^{\dagger}$
\footnote{Also at Imperial College, London and Lebedev Institute, Moscow}\\
\vskip 0.4truecm
$^{*}$
{\it Max-Planck-Institut f\"ur Gravitationsphysik,
Albert-Einstein-Institut, \\
Am M\"uhlenberg 1, D-14476 Golm, Germany}\\
\vskip .2truecm 
$^{\dagger}$ {\it Department of Physics,
The Ohio State University,\\
Columbus, OH 43210-1106, USA}\\}
\end{center}
\vskip 0.2truecm \noindent\centerline{\bf Abstract}
We show that a low-energy  action for  
massless fluctuations around a tachyonic soliton background 
representing a codimension one D-brane coincides with the 
 Dirac-Born-Infeld  action. The  scalar modes which 
describe transverse oscillations of the D-brane
are translational collective coordinates of the soliton.
The appearance of the DBI action is a universal 
feature  independent of details of a tachyon effective action, 
provided it has the structure implied by the open 
string sigma model partition function.

\newpage

\def \ci{\cite}
\newcommand{\rf}[1]{(\ref{#1})}

\renewcommand{\thefootnote}{\arabic{footnote}}
\setcounter{footnote}{0}

\section{Introduction}
In the original perturbative 
 string-theory  description 
 D-branes \cite{DLP}  are specified by boundary conditions
on the open strings ending on them. Their
 collective coordinates 
are identified with massless modes of the 
 open strings, and the ``acceleration-independent"
 part of their action -- the DBI action \ci{L} -- 
 can be derived 
 (using T-duality) 
  as a reduction of Born-Infeld action
 (directly  from the open string sigma model
 partition function  \ci{FT} 
 or from conformal invariance condition \ci{AB}, 
 see  \ci{L,Bac,TT}).

Supersymmetric 
D-branes appear also in another guise 
as black-hole type solitons of $D=10$ type II 
supergravity 
\ci{HS,DL} which carry Ramond-Ramond charges \ci{P}.
Their  collective coordinates, 
determined by the 
 massless fluctuation modes in these
backgrounds,
 can  be related to 
 the parameters of spontaneously broken gauge 
symmetries \ci{Cal}. With
an  appropriate non-linear  parametrization 
of the supergravity fields in terms of the collective
coordinates,   one should be able to derive the 
corresponding DBI actions directly from the  
type II supergravity action.\foot{To a large extent 
this  should  follow
essentially from reparametrization 
 invariance, 
implying that the action of a boosted 11-d Schwarzschild 
black hole should be $M\int dt  \sqrt{1-v^2}$, and
 T- and S- dualities of type II supergravities.}

Recently, a  new, third, 
 description  of D-branes as 
 solitons was suggested \cite{Sen,SS}.
Arguments supporting the proposal 
that D-branes can be interpreted as tachyonic solitons
were given 
using a variety of approaches:
 boundary conformal field theory (see, e.g.,  \ci{HKM}),
   Witten's open string field theory \ci{Wsft}
(see, e.g., \ci{SFT}),  non-commutative  
 field theory obtained in large $B_{\m\n}$ limit 
(see, e.g., \ci{HLK}) and  simple 
low-energy effective Lagrangian models
with specific tachyon potentials \ci{MZ1,MZ2,MZ3}.

Remarkably, the 
 latter models were shown to follow 
from the  boundary string field theory (BSFT) 
 \cite{Witten,Sch} in \ci{GS,KMM1,KMM2}.
The BSFT approach, consistently restricted to the lowest-level 
(renormalizable)  tachyon 
and  vector  couplings, is essentially  the same 
as the  off-shell  sigma model approach \ci{FT1,FT,T}
being based on the disc partition function 
of the open string sigma model.
The partition function encodes the information not only  about 
the beta-functions but also about the field space 
metric which relates them
to field equations of motion, and  thus 
allows  to smoothly interpolate between the standard perturbative
tachyon
vacuum and a new non-trivial  vacuum at the minimum 
of the tachyon potential.\foot{A simple 
consequence of the sigma model
approach to the tachyon condensation 
is the background independence of
the tachyon potentials discussed in the 
framework of Witten's open string field
theory in \cite{Seni}:  the zero mode
of the tachyon field which is the only argument 
of the potential 
 does not feel any closed string background.}
One finds that 
there are solitonic solutions corresponding to D-branes of
 lower dimensions 
\cite{MZ1,MZ2},
and that the descent relations between  D-brane tensions
hold exactly \cite{KMM1,KMM2}.
 These results 
have been generalized to include a  background 
 gauge field
\cite{LIW,C,O,A,GS1,T}.\foot{Since a constant 
field strength $F_{\mu\nu}$ enters the
actions in
the same way as a constant $B_{\m\n}$-field, 
the sigma model approach 
should  provide a
natural explanation for  some 
of the results for noncommutative tachyon
condensation 
 in \cite{HLK}.}
   
  {}From the boundary sigma model point of view, one   starts  with a  
 conformal theory with $d$ Neumann boundary conditions in the UV  
and adds relevant (tachyon) perturbations driving the theory 
to an IR fixed point that corresponds to a new (stable or unstable) 
vacuum with $(d-1-p)$ Dirichlet boundary conditions. The IR fixed 
point is then interpreted as a closed string vacuum with a Dp-brane.     
Given that  the tension of a Dp-brane is correctly 
reproduced \cite{KMM1,KMM2},
a further crucial test  
is to find  the spectrum and the effective action 
for light modes on 
 a Dp-brane obtained as a result of tachyon condensation. 
 
The aim of the present paper is to show that 
(the ``acceleration-independent" part of) the action 
for the massless scalar modes 
 $\Phi$ on  the tachyonic soliton and the massless
$U(1)$ vector field $A$ is indeed the standard  DBI action. 

In the context of the Sen's  proposal to describe D-branes 
as open string field theory solitons  
the massless  modes  representing  transverse D-brane  fluctuations 
appear as collective coordinates. 
The presence of these  massless scalars is 
a general phenomenon independent of details of an 
effective  field theory action, and is  a consequence 
of  spontaneous breaking of the
translational symmetry \ci{CC}. 
The existence of a massless vector  mode  on the soliton 
 is related to 
the fact that the tachyon is coupled to the (abelian) 
open string 
vector field in a non-minimal 
way, i.e.  only through the field strength $F_{\m\n}$.\footnote{Note that 
one   does not find massless vector modes and DBI actions
 for the collective coordinates for brane 
 solitons in familiar scalar-vector   field-theory   systems 
 where 
scalars are coupled to vector fields in a 
minimal way.}
The general structure 
 of the dependence of the effective action on $F_{\m\n}$
and its coupling to the tachyon $T$ is dictated
by the open string sigma model \ci{FT,AB,T}: 
 the action is an integral of 
  a product of the Born-Infeld Lagrangian 
for the vector field and some function of $T$ and its derivatives 
contracted  with  functions of  $F_{\m\n}$.
This structure is  
 the only  essential assumption one   needs  
to show that the   scalar and vector zero modes always 
combine into the standard DBI action, 
irrespective of all other details of the effective action.\foot{The precise
 value of the tension
 of the resulting 
Dp-brane  does depend of course on details of an 
effective string field theory action.
To reproduce the expected tension 
one should  compute the disc  partition 
at the proper conformal point, or minimize 
the ``trial actions" depending on a 
 finite number of parameters
 obtained  in the BSFT framework 
 in \cite{KMM1,KMM2}.} 

While  the universality of the DBI action holds irrespective of the 
presence of a tachyonic mode in the spectrum of fluctuations around 
the  soliton, the form of an action for the tachyonic 
mode\footnote{The
tachyonic mode is always present in  open bosonic string theory
context, and is 
absent on the kink solution describing condensation of an
unstable D9-brane tachyon to produce  a D8-brane, which
is a stable BPS-object in type IIA theory.}  does depend on 
details of an effective string field theory action (the  
action  for the tachyonic mode  
on D24-brane was shown \cite{MN} to have the same form as the 
one on the D25-brane). 

We should add that  while  finding the DBI action 
for the D-branes described by tachyonic solitons was 
expected -- after all, the required information is
contained in the open string sigma model which gives 
both, the effective action for the original brane
and the action for massless fluctuations of the solitonic brane -- 
to see how this derivation works in detail seems 
quite instructive. In particular,  this clarifies
how the brane collective coordinates should  appear 
in the string  world sheet action which may be relevant 
for off-shell aspects of D-branes
(cf. \ci{recoil}).

In Section 2 we shall 
present a  derivation of the DBI action from an 
effective  field theory action  with the structure implied by 
the open string sigma model. In Section 3 we shall 
discuss  how one can see the same 
directly in the world-sheet 
approach, identifying the boundary  couplings of the  D-brane 
collective coordinates.

\section{Effective field theory picture}

\subsection{Tachyon--vector  actions}

The string sigma model  approach to 
  field theory 
describing dynamics of an unstable D25-brane 
in bosonic string theory or 
a non-BPS D9-brane in type IIA string theory 
leads to an effective action for the tachyon 
and vector gauge fields 
of the following general form (for a  review, see \cite{T})
\footnote{In the units  we shall use $\a ' =2$, and our
normalizations for $T$ and $A_{\m}$ are given in Section 3.}
\bea
S = \int d^{d}x \sqrt{\det (\d_{\m\n} 
+ F_{\m\n})}\ {\cal L} (T, F, \pa T, \pa  F,
\pa \pa T, \ldots) .
\label{actiong}
\eea
The tachyon field  couples to the vector 
field through its  strength, i.e.   through the combinations
 (and their derivatives)\foot{The case when  the starting point is 
 a  Dp-brane with a non-zero  number of transverse dimensions
 represented by massless scalars  is obtained by ``dimensionally reducing" 
 $F_{\m\n}$ matrix.}  
 the 
\bea
G^{\m\n}=  \left( \frac{1}{1+F}\right)^{(\m\n)}=
\left( \frac{1}{1-F^2}\right)^{\m\n}\ , \ \ \ \ \ \ \ 
H^{\m\n}=  \left( \frac{1}{1+F} \right)^{[\m\n]}=
-\left( \frac{F}{1-F^2}\right)^{\m\n}\ , \label{mata}
\eea 
where $G^{\m\n}$ plays the role of  an effective metric
as seen by open string excitations  in 
the presence of a constant gauge field background \ci{AB}. 

We will be interested in 
deriving  an effective action for massless modes on
a soliton solution up to ``acceleration-dependent" (i.e. 
second and higher derivative) terms. 
For that reason we may restrict our consideration to 
the terms in $S$ which do not
depend on derivatives of $F_{\m\n}$.
 We  will study only codimension one D-branes, i.e.  
soliton solutions represented by  the tachyon field 
depending only  on one 
coordinate $x_1$ and the  vector  field having 
constant field strength $F_{\mu\nu}$.
As will be shown below,  such backgrounds
satisfy the equations of motion for
the vector field only if 
$F_{1i}=0$ ($i=2,...10$),  i.e.,  if the matrices in \rf{mata}
obey
$G^{1i}= H^{1i}=0$.

Since in deriving the effective action for massless modes we drop
their 
second and higher derivatives, none of the 
possible $H^{\m\n}$
dependent terms in  \rf{actiong}
contribute to the effective action,
i.e.  we are allowed to use 
an action which depends only on $G^{\m\n}$ and  $T$ and its
 derivatives of
any order, i.e.
\bea
S = \int d^d x \sqrt{\det (\d_{\m\n} 
+ F_{\m\n})}\ {\cal L} (G^{\m\n}(F), T, \pa T, \pa\pa T, \ldots )\ .
\label{action}
\eea
Though  any 
action of the form (\ref{action})  will 
lead to the DBI action for the  massless soliton modes, 
it is useful   to recall  several explicit 
 actions of that type 
   which were  previously discussed  in the
context of tachyon condensation.   
The  low-energy two-derivative effective  
action for a D25-brane in open bosonic string theory is \rf{action}
with  ${\cal L}$ given by  \cite{GS1,T} 
\bea
{\cal L}_{25}=T_{25} \ {\rm e}^{-T}\ (1+T) 
\bigl( 1+G^{\m\n}\pa_{\m}T \pa_{\n}T \bigr).
\label{D25}
\eea
The corresponding  expression  for  the  
 non-BPS D9-brane in type IIA theory is \cite{KMM2,T}
\bea
{\cal L}_9 = T_9\ {\rm e}^{-\frac14 T^2}\left( 
1 +   G^{\m\n}\pa_\m T\pa_\n T
+ { 1 \over 2}  \log ( \frac{4}{{\rm e}} )\  G^{\m\n}T^2 \pa_\m T 
\pa_\n T \right) .
\label{D9}
\eea
The  actions (\ref{D25}) and (\ref{D9})
are obtained  in special schemes chosen  to 
 reproduce the standard 
values of the tachyon masses around the perturbative  $T=0$ vacua.
The actions (\ref{D25}) and (\ref{D9}) 
(or similar model actions in  \cite{MZ1,MZ2})
have soliton solutions representing codimension one D-branes, but they 
do not reproduce the expected D-brane tensions.
 However, one can find 
``interpolating" actions which 
coincide with (\ref{D25}) and (\ref{D9}) in the 
two-derivative approximation and lead to 
the correct  D-brane tensions. 
In particular, for  a non-BPS D9-brane such  action 
 has   \cite{KMM2,A,T}
\bea
{\cal L}_9 = T_9\ {\rm e}^{-\frac14 T^2}\left( 
\sqrt{\pi}\frac{\G(1 + G^{\m\n}\pa_\m T\pa_\n T)}
{\G(\frac12 + G^{\m\n}\pa_\m T\pa_\n T)}
+ \log ( \frac{4}{{\rm e}})\ G^{\m\n} T \pa_\m\pa_\n T\right) .
\label{d9b}
\eea
Here the first term can be shown to coincide with the action 
found in \cite{KMM2,A}, and the scheme-dependent coefficient 
of the last 
 term  is chosen to reproduce  the 
correct value of the tachyon mass at 
$T=0$ \cite{T}. 

The action (\ref{action}),(\ref{d9b}) 
is given by   the open superstring partition function 
on a disc. 
In the sigma model approach
the tachyon coupling at  the boundary of the disc 
can be  represented as 
\bea 
T(x(\tau )) = T(x) + \xi^\m (\tau ) \pa_\m T(x) + 
\frac12 \xi^\m (\tau )\xi^\n (\tau )\pa_\m \pa_\n T(x) +
 \cdots, \label{texp}
 \eea
where $\tau$ is the angular coordinate on the boundary of the disc, 
and $ x(\tau)= x + \xi (\tau)$, \ \  $\int d\t \xi (\tau )=0$.
The derivatives of the tachyon field are then treated as 
independent, and the 
partition function is computed exactly in $\pa_\m T$, and up to 
the first order in $\pa_\m \pa_\n T$.

\subsection{Collective coordinates and DBI action}
We shall assume that the equations for $T $ and $A_\mu$
following from   (\ref{action}) 
have  a soliton solution  with 
\bea T=T(x_1)\ , \ \ \ \ \ \ 
F_{\mu\nu}=\const\ . \label{back}
\eea
Following Sen \cite{Sen}, 
we  shall identify the solution with a D(d-2)-brane.
For example, for  model two-derivative actions
  one finds  the kink 
$T(x) = u x_1$ \cite{MZ2,KMM2} in the case of the type IIA 
D9-brane, and the lump $T(x) = u x_1^2 $ 
\cite{MZ1,KMM1} in  the case of the bosonic  D25-brane.
These solutions represent exact conformal field theories in the  limit
$u\to\infty$  \cite{HKM,KMM1,KMM2}.

It is straightforward to see 
that for any solution for $T$ found  for $F_{\mu\nu}=0$ 
there is a corresponding solution  for  $F_{\mu\nu}=const$. 
Since the vector potential $A_{\mu}$ is a dynamical field, 
one is still  to check  that its  equation
is satisfied  for  the above  background.
To do that 
we  shall restrict ourselves to
the simplest case when  the action  can be put in the form 
\bea
S = \int d^d x \sqrt{\det (\d_{\m\n} 
+ F_{\m\n})}\ {\cal L} (G^{\m\n}(F)\pa_\m T\pa_\n T, T )\ .
\label{action1}
\eea
The explicit examples of the 
actions  considered above are all of that type.
Then varying (\ref{action1})  with respect to 
$A_{\mu}$ one gets (assuming \rf{back}) 
\bea
\pa_{\m}\Biggl[ (I+F)^{-1}_{[\m\n]}\  {\cal L} 
+  2 (I+F)^{-1}_{1[\mu} (I+F)^{-1}_{\n] 1} 
(\pa_1T)^2 {\cal L}'\Biggr]=0. 
\eea
Here ${\cal L}={\cal L}( K, T), \  K\equiv G^{\m\n}\pa_\m  T \pa_\n
T$ 
and  the derivative ${\cal L}' $ 
is with respect to the first argument $K$.
Since ${\cal L}$ depends only on $x_1$ 
this equation reduces to
\bea
\label{eqm}
 (I+F)^{-1}_{[\n1]} \ \pa_{1}\bigl[
{\cal L} - 2 G^{11}(\pa_1T)^2 {\cal L}'\bigr]
= 0\ . 
\eea
This equation is satisfied if
\bea
(I+F)^{-1}_{[\n1]}= \bigl( \frac{F}{1-F^2} \bigr)_{1\n}
=0\ , \eea 
i.e. if  $F_{1\n}=0$, {\it or}  if
\bea
{\cal L} - 2 G^{11}(\pa_1 T)^2 {\cal L}' = {\rm const}\, .
\eea
The latter  condition does not hold
  for a generic Lagrangian, 
so we will always require that $F_{1\n}=0$.
 Then 
the effective metric $G^{\m\n}$ simplifies,  so that 
\bea G^{11}=1,\ \ \ \quad G^{1i}=0. \label{onnn} \eea 
Note also that in  the axial gauge, $A_1(x_1,x_i)=0$, 
 the condition
$F_{1\n}=0$ implies that $A_i(x_i)$ does 
not depend on $x_1$,  as one would 
expect for a massless 
vector field on a D(d-2)-brane.

It follows from the translational invariance 
of the action (\ref{action})
that fluctuations around 
the soliton solution contain a zero mode $\Phi (x_i)$, which 
satisfies the equation of motion for a massless scalar 
in $d-1$ dimensions.
As usual (see, e.g., \cite{CC}),
 it  is  natural to identify this  zero mode 
with a collective coordinate
describing fluctuations of the D(d-2)-brane 
in the transverse $x_1$  direction. 
For small fluctuations and in the semiclassical approximation, 
the effective action
for the zero mode is obtained by substituting 
$T \to T(x_1) - \pa_1T(x_1)\Phi (x_i)$ 
into (\ref{action}), 
and integrating over $x_1$.
 To  describe arbitrary fluctuations in the semiclassical 
approximation one may then consider the standard  ansatz
$T \to T(x_1 -\Phi (x_i))$.  

However, in general, one should take into account
 the global $SO(d)_G$ 
invariance  of the action 
(\ref{action}) implied by  the open string sigma model. 
The subscript 
$G$ means that the  general linear  
transformations of the  $d$ 
coordinates  $x \to \Lambda x $  should 
 preserve the  constant  metric $G^{\m\n}$, i.e.
$$ \Lambda G \Lambda^T = G .$$
Assuming that $\Lambda$ has the form 
\bea 
\Lambda_{\m\n} = \frac{1}{\b}\tilde{\Lambda}_{\m\n},\quad \ \ 
\tilde{\Lambda}_{11}=1,\ \ \quad \tilde{\Lambda}_{1i} = -V_i \ , \eea 
we get 
\bea \b =\sqrt{ 1 + G^{ij}V_iV_j } \ . \eea 
Consider now the  following ansatz  for the 
tachyon  in terms of  the zero mode $\Phi (x_i)$
\bea
\label{tach}
&&T= T(y_1) \ , \ \  \ \ \ \ \ \
 y_1 \equiv 
\frac{x_1 - \Phi (x_i)}{\b (\pa \Phi, F )}\ , \\
&&\b (\pa \Phi, F ) =  
\sqrt{1 + G^{ij}(F) \pa_i\Phi\pa_j\Phi} \ . 
\eea
Note that if $\Phi (x_i)$ is 
of the form $\Phi (x_i)=V_i x^i$
(i.e. brane has  ``constant velocity"),  
then $y_1=(\Lambda x)_1$ and $\L$ belongs to the 
invariance group $SO(d)_G$. 

Taking into account 
that the  full tachyon field in \rf{tach} 
  depends only on  one combination $y_1$
of coordinates, we can  rewrite (\ref{action}) as follows
(ignoring  higher-derivative  $\pa^n\Phi ,\, n>1$ terms)
\bea
S= T_{d-2} 
\int d^{d-1}x\  \b (\pa \Phi, F  ) \sqrt{\det (\d_{\m\n} 
+ F_{\m\n})}\ ,
\label{action4}
\eea
where the integral over $y_1$
\bea
T_{d-2} \equiv  \int dy_1 {\cal L} ( T(y_1), \pa_1 T(y_1),
\pa_1 \pa_1 T(y_1), \ldots )\
\label{ion}
\eea
 does not depend on the field strength\footnote{As  follows from 
 the sigma model considerations,
   $G^{\m\n}$ can be 
 contracted only with the tachyon derivative terms.
  Thus only  $G^{11}$  could 
appear in (\ref{action4}), but it is equal to 1 according to 
\rf{onnn}.} 
and  thus 
defines the D(d-2)-brane tension.

Taking into account that (for $F_{i1}=0$) 
\bea
\b (\pa\Phi, F ) \sqrt{\det (\d_{\m\n} 
+ F_{\m\n})}= \sqrt{\det (\d_{ij} + F_{ij}
 + \pa_i\Phi\pa_j\Phi )}\  ,
\label{dbi}
\eea
we arrive at the standard DBI action for a D(d-2)-brane
\bea
S=T_{d-2} \int d^{d-1}x \sqrt{\det (\d_{ij} + F_{ij}
 + \pa_i\Phi\pa_j\Phi )}\  .
\label{dbi2}
\eea
 To get  the DBI action it was essential to
introduce the collective coordinate dependence as
 in (\ref{tach}). 
 Indeed, the above  derivation  is a generalization of the argument
used to show that  a point-like  soliton of a relativistically 
invariant (e.g. 2-d) scalar field 
 theory  is described by the standard  particle 
action $M \int dt \sqrt {1- v^2} $: one starts with the static 
solution and applies a boost $ x \to { x - v t \over \sqrt {1 - v^2} }  $. 
The  ansatz (\ref{tach}) of course  reduces to 
$T(x_1) - \pa_1T(x_1)\Phi (x_i)$ for small fluctuations, 
but the  real reason  for the specific  non-linear structure 
of  (\ref{tach}) is to ensure  the decoupling
of massive solitonic modes from the massless mode, 
so that  there is no term linear in the massive modes in the action 
(in the approximation where 
we neglect second and higher derivatives of $\Phi$).
Thus one can consistently put all massive modes to zero, and consider only 
the 
$\Phi$-dependent terms.\foot{This is somewhat 
similar  to  what happens 
in consistent Kaluza-Klein  reductions of compactified gravity models.}   
If one  would use the  standard  form of the 
collective coordinate dependence  $T=T(x_1 - \Phi (x_i))$ 
then to recover
the DBI action one would have to take into account 
the massive mode contributions
to the low-energy effective action and  at the end 
to make a proper non-linear 
 redefinition of the collective coordinate $\Phi$.


\section{World-sheet picture}
Here  we shall relate the discussion  of the 
DBI action from the effective field theory point of view 
in the previous section  to the 
world-sheet sigma model considerations, 
with the aim to identify  the  boundary couplings 
corresponding to the tachyonic soliton collective coordinate.

\subsection{Standard boundary couplings}

The starting point in the usual perturbative description 
of D-branes in flat space 
 is an exact free 2-d  conformal field 
 theory in the bulk with 
$p+1$ Neumann and $d-p-1$ Dirichlet boundary conditions. 
The marginal  boundary interactions are then represented by 
the boundary sigma model action \cite{L,DLP}\foot{For the sake 
of clarity  here we shall 
 consider only the bosonic string theory
 but will  ignore  tachyonic coupling on the resulting  brane.}
\bea
\label{Sbr}
I_{\pa\S}=\frac{1}{4\pi}\int_{\pa \S}d\tau \biggl[
\sum_{i=1}^{p+1} A_{i}(x^1,...,x^{p+1})\dot{x}^{i}+
\sum_{a=p+2}^d \Phi_{a}(x^1,...,x^{p+1})\pa_n {x}^{a}\biggr] \, , 
\eea  
where the couplings 
$A_i$ and $\Phi_a$ are the gauge field and the scalars on 
the world-volume of the $Dp$-brane, respectively ($\pa_n$
is the normal  derivative to the boundary $\pa \S$). 
 The open string sigma model 
 partition function on the disc is then the DBI
 action  for $A_i$ and $\Phi_a$ (ignoring terms
 higher than first derivative in the fields).

On the other hand,  to describe  tachyon condensation 
from a space-filling brane to a 
lower-dimensional brane  one should  start
 with the  following boundary  theory  
\bea
\label{Sbnonr}
I_{\pa\S}=\frac{1}{4\pi}\int_{\pa \S}d\tau \bigl[T(x)+A_{\mu}(x)\dot
{x}^{\mu}\bigr] \ . 
\eea  
In the presence of this  interaction the standard Neumann  boundary
conditions for the  open string coordinate
 fields $x^{\mu}$ are modified 
to 
\bea
\pa \S: \ \ \ \ \ \ \ \ \pa_n x_{\mu}+F_{\m\n}\dot{x}^\n +\pa_{\m}T=0\ . 
\label{main}
\eea  
 A particularly simple case for which 
the open bosonic world-sheet theory remains solvable is obtained by taking  
$A_{\mu}$ to have a constant field strength and switching on
the  tachyon  with a 
 quadratic profile in one direction, i.e.  
$T(x)=a+\frac{1}{2}u\,x^2_1$. In this  case the boundary conditions 
are 
\bea
\pa \S: \ \ \ \ \  \pa_n x_{1}+F_{1i}\dot{x}_i+u\,x_1=0,~~~~~~~~~
\pa_n x_{i}+F_{i1}\dot{x}_1+F_{ij}\dot{x}_j=0 \, . \label{bir}
\eea 
For  $u=0$ these 
 are the usual Neumann conditions modified by the 
presence of the constant strength $F_{\mu\nu}$ (and thus the 
 corresponding
partition function  is the BI action for $F_{\mu\nu}$ with 
$\mu =1,\ldots, d$).
  On the other hand, when the  tachyon condenses 
into  the vacuum $a,u\to \infty$, the field $x_1$ becomes
 constrained to vanish at the boundary, i.e.
is subject to 
  the  Dirichlet boundary condition  ($x_1|_{\pa\S}=0$). This 
leads to a complete decoupling of $F_{1i}$ components 
in \rf{Sbnonr},\rf{bir}.  
Thus, in the IR fixed  point one gets the BI action for $F_{ij}$
components  only. 

 Following this approach,  it appears 
that we are  missing the  dependence
on the massless scalar fields describing 
the transverse motion of the brane.
To account for them, one  should 
understand how the normal derivative coupling  of the type
introduced from the beginning  in the standard D-brane description 
 \rf{Sbr}  is effectively induced 
in the process of  tachyon  condensation.
This is what we are going to explain below.

\subsection{ Generalized boundary sigma model  and collective coordinate
coupling     }

{}From the sigma model viewpoint, 
the role of the tachyonic field is to control 
a number of ``free"  (Neumann) 
space-time dimensions,  and in this respect 
the condensation may  be compared with 
dimensional  reduction. However,   
in the process of reduction  
 the BI action in $p+1$ dimensions
does lead to 
 the  DBI action in $p$ dimensions  with $A_1\to \Phi$.
 The procedure of dimensional 
 reduction  is of course  formally related
  to T-duality transformation 
 along  one world-volume dimension $x_1$ (implying that $A_1$
 becomes a scalar field describing the transverse fluctuations).
 
In the  world-sheet  description, 
the T-duality transformation  simply exchanges the    
normal and tangential  derivatives in the boundary vertex 
operators \ci{DLP}. 
This suggests that 
in order to make contact with the D-brane description
based on   (\ref{Sbr})
one  may try   to  generalize   the boundary  action 
(\ref{Sbnonr})  by adding  from the beginning an  
additional
normal derivative coupling (with a new coefficient 
function $V_{\m}$)\foot{Such sigma model 
and its renormalization was discussed in detail in \ci{KV}
and references there.} 
\bea
\label{Sbnonr1}
I_{\pa\S}=\frac{1}{4\pi}\int_{\pa \S}d{\tau} \biggl[T(x)+
A_{\m}(x)\dot{x}^{\m}+V_{\m}(x)\pa_{n}x^{\m}\biggr] \, . \label{vvv}
\eea
Adding $V_{\m}$ term  may seem 
irrelevant  -- in 
the standard perturbative vacuum where all $x^\m$ 
are subject to 
 the Neumann boundary conditions this term 
 decouples: computing  correlation functions or 
the disc  partition function in perturbation theory
in powers of couplings  one finds that they do not
 depend on $V_\m$.\foot{There  exists  a (point-splitting)
 regularization
preserving the Neumann boundary conditions,
 and, therefore, any correlation
function involving a  normal-derivative term vanishes.
Note  also that  lifted to the bulk of the world sheet, 
this coupling becomes $\int_{\S}\pa^a( V_{\m}(x)\pa_{a}x^{\m})$, 
and thus (modulo a term proportional to the $x^\m$ equation of motion)
redefines the target space metric  by a diffeomorphism-type 
term.
}

More precisely,  one 
 should determine  the boundary conditions 
dynamically, 
 by minimizing the total string action $I=I_{\S} + I_{\pa\S}$. 
Then (depending on a  calculation procedure)
the normal-derivative term  may  contribute 
to the partition
function,  but this  still  does not imply 
 the appearance 
of a new physical  degree of freedom associated with $V_\mu$.  
Indeed,  one may interpret  such normal derivative couplings 
as additional  pure gauge  modes 
of open string theory,  which  can be removed 
by  a gauge choice. 

Indeed, the general form of the 
 boundary conditions  should be 
\bea
\label{nbc}
\pa \S: \ \ \ \ \ \ 
\pa_{n}x_{\m} + N_{\m\n}\dot{x}^{\n}+ N_{\m} = 0,
\eea
where   $N_{\m\n}$ and $N_\m$  should be determined 
from the requirement that 
the variation of the total string action ${ 1 \over 8 \pi} \int_\S 
(\pa
x)^2 + I_{\pa\S}$ vanishes.
One finds then the following equations\foot{These
boundary conditions differ from the ones used in \cite{KV}
and references there.}
\bea
&&N_\m =\pa_\mu \tilde {T}\ , \ \ \ \ \   \tilde {T}\equiv
T - N_\n V^\n  \ , \\
&&N_{\m\n} =\tilde { F}_{\m\n}\ , 
\ \ \ \ \ \ \tilde { F}_{\m\n}\equiv 
 F_{\m\n}-\pa_\m (N_{\rho\n} V^\rho )+\pa_\n (N_{\rho\m} V^\rho )
 \, .
\eea
It is then easy to see    (cf. \rf{nbc},\rf{main})   that 
one can replace \rf{Sbnonr1}  by  the  usual 
boundary term  \rf{Sbnonr} without $V_\m$ but with transformed 
tachyon  
$T\to \tilde {T}$ and vector $A_\m \to \tilde { A}_{\m}$ fields: 
\bea
T=( 1 + V^\m \pa_\m ) \tilde {T}\ , \ \ \ \ \ \ \ \ \ \
A_\m = \tilde {A}_{\m} - \tilde {F}_{\m\n}V^\n \, . 
\eea
Thus adding the normal-derivative term 
amounts simply to a redefinition of the 
 original  open string sigma model couplings  in (\ref{Sbnonr}).

The central  point, however, is that 
 in the presence of a nontrivial
tachyon condensate  the normal derivative coupling
becomes relevant --  it provides an adequate 
description  of dynamics of the soliton  translational zero modes, 
allowing one  to recover the usual 
D-brane description in the IR.

Assuming that there is  
a tachyonic condensate breaking
translational invariance in $x_1$-direction,
 and that $V_1$ depends only on
$x_i$, one sees that 
$V_1$ can be identified with 
the translational collective coordinate $\Phi$.
Indeed, for small $V_1$ 
\bea
\tilde {T}\approx  T -  V_1 \pa_1 T\approx  T(x_1 - V_1) ,\quad\ \ \ \ 
\tilde {A_i}\approx A_i -  V_1 \pa_1 A_i \approx  A_i(x_1 -  V_1),
\label{tv}
\eea
which is the standard collective coordinate dependence. 
Another argument for this  identification can be  given  by  
analyzing the boundary conditions (\ref{nbc}) in the IR fixed point 
$a, u\to\infty$ for the  quadratic tachyon coupling 
$T(x_1)= a + \frac12 u x_1^2$. 
At this  point the boundary condition (\ref{nbc}) reduces to the 
following Dirichlet condition for $x_1$ 
\bea
\label{D}
\pa \S: \ \ \ \ \ \ \ \ \    x_1=\Phi(x_i) \, ,\ \ \ \ \ \ 
\Phi\equiv V_1 (x_i) \ , 
\eea
and  the modified Neumann condition for $x_i$  (cf. \rf{bir}) 
\bea
\label{R1}
\pa \S: \ \ \ \ \ \ \ \ \pa_n x_i+\pa_i\Phi\pa_n x_1+F_{ij}\dot{x}_{j}=0\, .
\eea 
The boundary conditions (\ref{D}) and (\ref{R1})  define the mixed 
Dirichlet-Neumann sigma-model.
 The vanishing of the one-loop $\beta$-functions
for this model is known \cite{L}
 to be equivalent to the equations of motion 
for space-time fields $A_i$ and $\Phi$ following from DBI. 

One  may  ask then  why in order  
to derive the DBI action   in Section 2 we needed to 
use the non-linear  ansatz for the tachyon (\ref{tach})
instead of the 
transformed tachyon and gauge fields in  (\ref{tv}).
The redefined  tachyon coupling $\tilde T$ 
 with the $V_1= \Phi$ -dependence as in (\ref{tach}) 
can be obtained by  starting with the  boundary action 
\rf{Sbnonr1}
with 
normal-derivative couplings of an appropriate  
non-linear form $f(V,\pa_{n}x,\dot x)$.
This  would modify  the boundary conditions (\ref{nbc}) 
and the $V_\m$-dependence of the transformed tachyon and vector fields.
{}From the sigma model point of view the  
difference between the  linearized 
 collective coordinate dependence
 in \rf{tv}  and the  non-linear 
 one in  (\ref{tach})  should be in contact terms only.
 Such terms  are crucial for implementing the symmetries of 
 string  theory in  a manifest way (see, e.g., \ci{CONT}), 
 but they do not 
 contribute to the (scheme-independent)
 one-loop $\beta$-functions.
 This explains why   one can
reproduce correct $\beta$-functions  corresponding to the 
DBI action  in the approach of \cite{L} 
using the linear expressions
 \rf{tv}. 
To derive the DBI action  as the open string sigma model 
partition function (which is sensitive to a  choice of a scheme
preserving underlying symmetries of the theory)
one  needs, however, to use 
the non-linear ansatz  (\ref{tach}) 
 for the collective coordinate dependence, 
 substituting it into the tachyon expansion 
 \rf{texp} and integrating over the fluctuations 
  $\xi^\mu$.

\vskip 0.3cm
\noindent
{\bf Acknowledgements}

\noindent
G.A. was supported by the DFG and by the European Commission 
RTN programme HPRN-CT-2000-00131 in which G.A. and S.T. 
are associated to U. Bonn, 
and in part by RFBI grant N99-01-00166 and by INTAS-99-1782.
S.T. also acknowledges support from GIF, 
the German-Israeli foundation
for Scientific Research.
S.F. and A.A.T are  supported 
 by the U.S. Department of Energy under grant
No. DE-FG02-91ER-40690. S.F. is also supported 
  in part by RFBI grant N99-01-00190,  and  
 A.A.T. --  by the 
INTAS project 991590
and PPARC SPG grant  PPA/G/S/1998/00613.

\newpage 



\end{document}